# Implications of kHz Quasi–Periodic Brightness Oscillations in X–Ray Binaries for Neutron Star Structure

Arun V. Thampan[1*], Dipankar Bhattacharya[2†], and Bhaskar Datta[1,2‡]

[1] *Indian Institute of Astrophysics, Bangalore 560 034, India.*
[2] *Raman Research Institute, Bangalore 560 080, India.*

**ABSTRACT**
Kilohertz Quasi-Periodic Oscillations(QPOs) in low–mass X–ray binaries (LMXBs) may represent the orbital frequencies of innermost Keplerian orbits around accreting neutron stars. Attempts have recently been made to derive constraints on the mass and the equation of state of the neutron star, by identifying the highest observed QPO frequency with the Keplerian frequency at the marginally stable orbit given by the general theory of relativity. These estimates have either neglected the effect of the neutron star rotation or used an approximate treatment of rotation in general relativity. We rederive these constraints using a fully general relativistic formalism including the effect of rapid rotation. We also present constraints corresponding to the case where the innermost stable orbit touches the stellar surface.

## 1 INTRODUCTION

The recent discovery of kilohertz quasi–periodic brightness oscillations (kHz QPOs) in certain X–ray binaries (see van der Klis 1997 for a recent review) have led to suggestions that these are determined by the Keplerian orbital frequency near the marginally stable orbit predicted by general relativity. Such suggestions have led to attempts to derive constraints on neutron star structure (and possibly also the equation of state) using the kHz QPO data (Kaaret, Ford & Chen 1997; Zhang, Strohmayer & Swank 1997a; Kluźniak 1998). It has been assumed in these calculations that the kHz QPOs are generated in the innermost stable orbit and that the neutron star lies within this orbit. These attempts have either neglected the effect of neutron star rotation, or relied on the use of an approximate treatment of rotation in general relativity. Since rotation can be important for neutron stars in accretion driven old X–ray binaries (Bhattacharya & van den Heuvel 1991), a fully general relativistic treatment of the rotation, in deriving constraints on the neutron star structure using kHz QPO data, is imperative.

In this paper, we explore the consequences of the association of kHz QPO with Keplerian frequencies at the inner edge of the accretion disk, using a fully general relativistic formalism appropriate for a rotational space–time. Unlike the previous calculations, we do not make the restrictive assumption that the radius of the neutron star is less than that of the marginally stable orbit. In general, the inner edge of the accretion disk need not always be coincident with the marginally stable orbit ($r_{ms}$), but can be located anywhere outside this radius. If the radius of the neutron star is greater than $r_{ms}$, the innermost possible orbit will be located at the surface of the star. Since the structure and $r_{ms}$ for a rotating neutron star depend on two independent parameters, namely, the central density ($\rho_c$) and the spin frequency ($\nu_S$), a range of values of ($\rho_c$, $\nu_S$) will exist that will allow solutions for a Keplerian frequency corresponding to a specific kHz frequency. We discuss here two branches of these solutions: (i) the case where the inner edge of the accretion disk is identified with $r_{ms}$ (Solution I) and (ii) the case where the disk extends to the surface of the star (Solution II). The discussions so far in the literature are confined to cases akin to Solution I. This is for two reasons: (a) a constant difference between the kHz frequency peaks and its near equality to the frequency of the QPO peak observed in X–ray bursts are taken to imply that the difference frequency represents that of the rotation of the neutron star. The highest QPO peak is taken to represent Keplerian frequency at the innermost marginally stable orbit. Since the frequency of this peak is higher than the neutron star rotation frequency, $r_{ms}$ is taken to be greater than the radius of the neutron star (Kluźniak 1998) and (b) the highest frequency peak is ob-served to be a near constant for different sources and if this peak were to originate at the surface of the neutron star, it might be difficult to explain the constancy despite the different magnetic fields and accretion rates that these sources may possess (Zhang et al. 1997a). The argument (a) above assumes that the accretion disk touches the neutron star surface, only if it spins at the centrifugal mass-shed limit (in which case, spin frequency ≈ equatorial Keplerian frequency at the stellar surface). However, since factors such as instabilities driven by gravitational radiation reaction (Wagoner 1984; Andersson, Kokkotas & Schutz 1998) may limit the spin frequency of the neutron star, there is no compelling reason to believe that this assumption is generally valid. The arguments in favour of solution I remain source-specific, and a main reason for the popularity of this solution is that features relating to the accretion flow and X-ray emission from the boundary layer have been computed only for this case (Kluźniak and Wilson 1991; Hanawa 1991). Until similar detailed computations are performed for other situations (e.g. solution II), it is not obvious that solution I should always be the preferred scenario for kHz QPOs.

[*] e-mail: arun@iiap.ernet.in
[†] e-mail: dipankar@rri.ernet.in
[‡] e-mail: datta@iiap.ernet.in

## 2 CALCULATIONS AND RESULTS

We look for possible configurations of rotating neutron stars in general relativity, whose innermost "allowed" orbits possess a QPO frequency of 1220 Hz observed in the source 4U 1636-53 (Zhang et al. 1997b). This is the maximum value of the highest QPO frequency observed so far. Our calculations are performed for a representative choice of neutron star equation of state (EOS) models, with a view to obtaining a broad general conclusion.

There can be two possibilities for the innermost marginally stable circular orbits: (i) $r_{ms} > R$, (ii) $r_{ms} < R$ where $R$ is the radius of the neutron star. For the case (ii), the innermost "allowed" orbit will be located at the surface of the neutron star i.e. $r_{iao} = R$ and for case (i) $r_{iao} = r_{ms}$. As already mentioned, we designate (i) as Solution I and (ii) as Solution II. It may be noted here that the configuration for which the radius of the innermost stable orbit is just equal to the equatorial radius of the neutron star defines the boundary between Solution I and Solution II.

The equilibrium sequences of rapidly rotating neutron stars in general relativity and the corresponding values of $r_{ms}$ used to derive the results in this paper, have been calculated using the formalism reported in Datta, Thampan & Bombaci (1998). We have chosen the following EOS models for neutron star interiors: (i) Bethe & Johnson (1974) model V, (ii) Walecka (1974), (iii) Wiringa, Fiks & Fabrocini (1988) model UV14 + UVII (iv) Sahu, Basu & Datta (1993) (v) Baldo, Bombaci & Burgio (1997) model BBB2, (vi) Bombaci (1995) model BPAL12. In order to conform to the notation used by Arnett & Bowers (1977), we refer EOS models (i) as (C) and (ii) as (N). The EOS (iii)–(vi) are referred to in the text as: (UU), (SBD), (BBB2) and (BPAL12) respectively. Of these, (N) and (SBD) are stiff EOS, (UU), (C) and (BBB2) are intermediate in stiffness and (BPAL12) is a soft EOS. For each of these, we have calculated the Keplerian frequency ($\nu_K$) of a test particle in the innermost "allowed" orbit ($r_{iao}$) for a range of central densities and rotation rates of the neutron star.

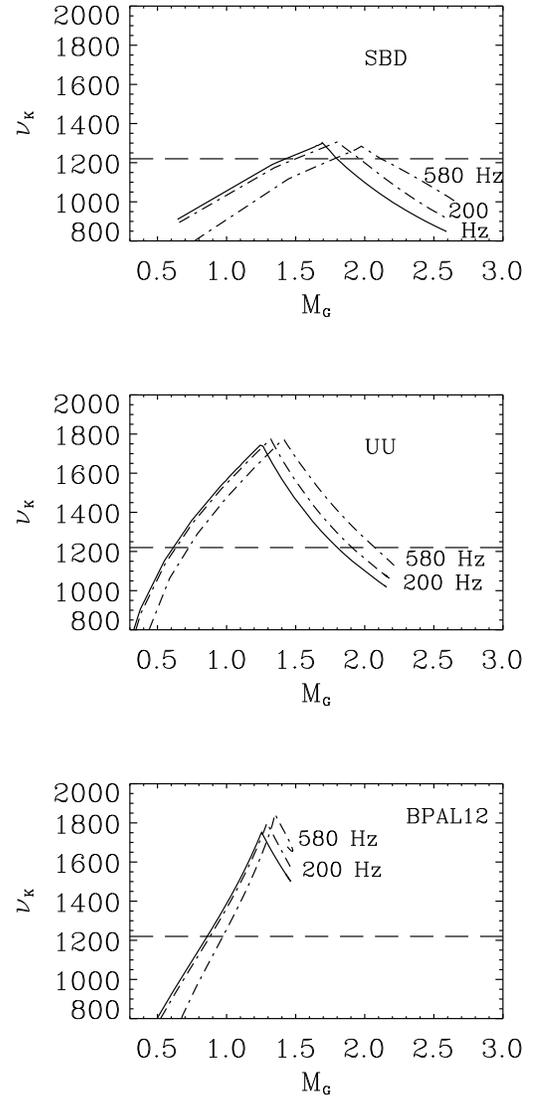

**Figure 1.** The Keplerian frequency $\nu_K$ corresponding to the innermost "allowed" orbit as a function of the gravitational mass $M_G$ of the neutron star. The three curves (one solid and two dot–dashed) are respectively for three values of neutron star spin frequency $\nu_S$ namely 0, 200 and 580 Hz (the last rotation rate inferred from 4U 1636–53 by Zhang et al. 1997b). The horizontal dashed line corresponds to $\nu_K = 1220$ Hz, the highest frequency QPO observed to date from the X–ray source 4U 1636–53. For all portions of the curves above the horizontal dashed line, 1220 Hz is obtained at some position exterior to the innermost "allowed" orbit. The values of $M_G$ at intersection points of the curves provide the range of masses that admit frequencies of 1220 Hz.

In Fig. 1 we have illustrated the dependence of the Keplerian frequency ($\nu_K$) on the gravitational mass ($M_G$) of the accreting rotating neutron star for three chosen values of its spin frequency ($\nu_S$): 0, 200 and 580 Hz. In this figure, we have considered only three EOS models: (SBD), (UU) and (BPAL12), which are respectively, stiff, moderate and soft, in order to illustrate the dependence of $\nu_K$ on the EOS.

**Table 1.** Range of masses corresponding to Solution I (intersection of the falling branch of curves in Fig. 1) for various rotation rates of the star for $\nu_K = 1220$ Hz. Only those EOS models which allow Solution I are displayed here. Listed in this table from left to right are: the EOS model, central density $\rho_c$, neutron star spin frequency $\nu_S$, baryonic mass $M_0$, gravitational mass $M_G$, equatorial radius $R$, specific angular momentum $j$ respectively of the neutron star, radius of the innermost marginally stable orbit $r_{ms,\mathrm{HT}}$ as calculated using HT formalism and radius of the innermost "allowed" orbit $r_{iao}$. In this case, $r_{iao}$ is equal to the radius of the marginally stable orbit dictated by general relativity. Only very slowly rotating configurations ($\nu_S < 100$ Hz) admit $\nu_K = 1220$ Hz for EOS model (BBB2) and hence there is only one entry against it. The numbers following the letter $E$ in the second column represent powers of ten.

| EOS | $\rho_c$ (g cm$^{-3}$) | $\nu_S$ (Hz) | $M_0$ (M$_\odot$) | $M_G$ (M$_\odot$) | $R$ (km) | $r_{iao}$ (=$r_{ms}$) (km) | $j$ | $r_{ms,\mathrm{HT}}$ (km) | $\nu_{K,\mathrm{HT}}$ (Hz) |
|---|---|---|---|---|---|---|---|---|---|
| (N) | 9.36E+14 | 0.0 | 2.252 | 1.802 | 12.315 | 15.964 | 0.000 | 15.963 | 1220.2 |
|  | 1.15E+15 | 440.7 | 2.687 | 2.079 | 12.336 | 16.612 | 0.205 | 16.080 | 1482.5 |
|  | 1.25E+15 | 580.0 | 2.833 | 2.169 | 12.344 | 16.813 | 0.287 | 15.968 | 1586.6 |
|  | 1.60E+15 | 825.0 | 3.118 | 2.336 | 12.146 | 17.167 | 0.370 | 15.848 | 1756.3 |
| (UU) | 1.37E+15 | 0.0 | 2.093 | 1.802 | 10.934 | 15.965 | 0.000 | 15.963 | 1220.2 |
|  | 1.50E+15 | 216.1 | 2.255 | 1.917 | 10.852 | 16.235 | 0.084 | 16.140 | 1313.0 |
|  | 1.85E+15 | 580.0 | 2.543 | 2.115 | 10.634 | 16.682 | 0.215 | 16.285 | 1477.5 |
|  | 2.23E+15 | 746.6 | 2.679 | 2.206 | 10.357 | 16.874 | 0.266 | 16.362 | 1543.5 |
| (SBD) | 4.81E+14 | 0.0 | 1.999 | 1.799 | 15.063 | 15.945 | 0.000 | 15.936 | 1222.3 |
|  | 5.10E+14 | 132.1 | 2.152 | 1.916 | 15.172 | 16.241 | 0.090 | 16.069 | 1326.6 |
|  | 5.20E+14 | 196.6 | 2.204 | 1.959 | 15.252 | 16.349 | 0.134 | 15.950 | 1395.4 |
|  | 5.35E+14 | 580.0 | 2.418 | 2.142 | 16.185 | 16.867 | 0.461 | 13.217 | 2289.5 |
| (BBB2) | 1.91E+15 | 0.0 | 2.093 | 1.802 | 10.378 | 15.965 | 0.000 | 15.963 | 1220.2 |

The innermost "allowed" stable orbit will be located at the neutron star surface for lower values of the mass (the rising branch of the curve in Fig. 1 – Solution II) and at the marginally stable orbit for higher values of the mass (the falling branch in Fig. 1 – Solution I). For a specific value of $\nu_K$ (taken here to be equal to 1220 Hz corresponding to the highest QPO frequency observed to date in the X–ray source 4U 1636–53), Fig. 1 then provides the mass of the neutron star (for a given $\nu_S$ and a given EOS): the intersection of the curves with the horizontal dashed line defines the neutron star configuration which possesses a Keplerian frequency of 1220 Hz at the innermost "allowed" stable orbit. It can be noticed that the peak heights of the curves above the horizontal dashed line decreases for increasing stiffness of the EOS. It is interesting to see that for model (SBD) (a very stiff EOS), the peak value of $\nu_K$ (1259 Hz in this case) for $\nu_S = 580$ Hz is very close to 1220 Hz. Stiff EOS models for which this peak lies below $\nu_K = 1220$ Hz line may then be disfavoured. From Fig. 1, further (QPO model dependent) constraints on the EOS and neutron star mass can be inferred as discussed below:

(a) If the highest frequency QPOs are produced only at the innermost marginally stable orbit (i.e if Solution II is to be disfavoured as suggested in Kluźniak 1998; Zhang et al 1997a; Kaaret et al. 1997), then the constraints obtained on the mass can be read off from the higher mass intersection points of the curves in Fig. 1. For soft EOS models, such an intersection may not obtain. This would then put a constraint against soft EOS models, a point noted in Kluźniak (1998). For intermediate and stiff EOS, the inferred mass values in this assumption for neutron star spin frequencies between 0 and 580 Hz, lies in the range (1.8–2.2) M$_\odot$. A lower mass limit of 1.8 M$_\odot$ using the approximate effects of rotation (Hartle–Thorne formalism) was also suggested by Kluźniak (1998). If the highest observed frequency in some QPO sources have lower values than 1220 Hz, the limits of this mass range will increase. For example, for a frequency of 1000 Hz, the EOS model (UU) (as well as BPAL12) will be disfavoured while the limit obtained from EOS SBD falls in the range (2.2–2.5) M$_\odot$.

(b) If the highest kHz QPO peaks are produced only at the innermost "allowed" orbit, then another set of constraints: lower limits on the neutron star mass, can be inferred from Fig. 1. For EOS models (BPAL12) and (UU), these limits are in the range (0.9–1.0) M$_\odot$ and (0.6–0.7) M$_\odot$ respectively. For EOS SBD, however, this range is (1.4–1.8) M$_\odot$, which is substantially higher than for the other EOS models. For lower values (than 1220 Hz) of the highest QPOs, the limits obtained from Solution II would allow even lower neutron star masses!

(c) For the non–restrictive assumption that the highest observed QPOs are produced at the inner edge of the accretion disk, which might be located *outside* the innermost "allowed" orbit, the only necessary condition imposed will be that $\nu_K \geq 1220$ Hz at the innermost "allowed" orbit. This implies a wider range (from the lower limit of the low mass intersection to the higher limit of the higher mass intersection) of inferred masses. In this case, we cannot rule out the validity of the soft EOS models, as the higher mass limit will be the maximum mass allowed by the EOS for the chosen value of $\nu_S$. It can also be seen that the constraints placed on the mass of the neutron star in this case are not as firm as in cases (a) and (b). For EOS models (BPAL12), (UU) and (SBD), the mass limits are in the ranges (0.9–

L4     *A.V. Thampan, D. Bhattacharya and B. Datta*

**Table 2.** Range of masses corresponding to Solution II (intersection of the rising branch of curves in Fig.1) for various rotation rates of the star for $\nu_K = 1220$ Hz. The listed quantities are the same as in the first eight columns of Table 1. In this case the radius of the star is larger than $r_{ms}$, and $r_{iao}$ is located at the stellar surface.

| EOS | $\rho_c$ (g cm$^{-3}$) | $\nu_S$ (Hz) | $M_0$ (M$_\odot$) | $M_G$ (M$_\odot$) | $R$ (km) | $r_{iao}$ (= R) (km) | $j$ |
|---|---|---|---|---|---|---|---|
| (C) | 6.81E+14 | 0.0 | 0.686 | 0.661 | 11.427 | 11.427 | 0.000 |
|     | 6.90E+14 | 198.9 | 0.706 | 0.679 | 11.520 | 11.520 | 0.147 |
|     | 7.00E+14 | 301.8 | 0.729 | 0.700 | 11.646 | 11.646 | 0.227 |
|     | 7.50E+14 | 580.0 | 0.847 | 0.809 | 12.268 | 12.268 | 0.468 |
|     | 1.06E+15 | 1129.6 | 1.503 | 1.384 | 14.695 | 14.695 | 0.983 |
|     | 1.16E+15 | 1209.3 | 1.657 | 1.512 | 15.090 | 15.090 | 1.000 |
| (N) | 5.37E+14 | 0.0 | 0.834 | 0.753 | 11.937 | 11.937 | 0.000 |
|     | 5.40E+14 | 122.4 | 0.849 | 0.765 | 11.986 | 11.986 | 0.094 |
|     | 5.50E+14 | 319.4 | 0.910 | 0.816 | 12.260 | 12.260 | 0.253 |
|     | 5.78E+14 | 580.0 | 1.092 | 0.968 | 13.041 | 13.041 | 0.503 |
|     | 6.50E+14 | 886.9 | 1.582 | 1.364 | 14.699 | 14.699 | 0.893 |
| (UU) | 6.30E+14 | 0.0 | 0.634 | 0.609 | 11.122 | 11.122 | 0.000 |
|      | 6.34E+14 | 142.0 | 0.645 | 0.619 | 11.174 | 11.174 | 0.107 |
|      | 6.50E+14 | 366.4 | 0.697 | 0.667 | 11.474 | 11.474 | 0.287 |
|      | 6.80E+14 | 580.0 | 0.800 | 0.762 | 12.036 | 12.036 | 0.486 |
|      | 8.80E+14 | 1106.0 | 1.527 | 1.399 | 14.788 | 14.788 | 1.071 |
|      | 9.70E+14 | 1210.7 | 1.802 | 1.627 | 15.485 | 15.485 | 1.117 |
| (SBD) | 4.11E+14 | 0.0 | 1.552 | 1.428 | 14.775 | 14.775 | 0.000 |
|       | 4.20E+14 | 190.0 | 1.632 | 1.496 | 14.958 | 14.958 | 0.148 |
|       | 4.25E+14 | 256.7 | 1.679 | 1.537 | 15.094 | 15.094 | 0.203 |
|       | 4.40E+14 | 401.1 | 1.826 | 1.661 | 15.511 | 15.511 | 0.331 |
|       | 4.55E+14 | 502.6 | 1.971 | 1.784 | 15.908 | 15.908 | 0.431 |
|       | 4.70E+14 | 580.0 | 2.109 | 1.900 | 16.270 | 16.270 | 0.509 |
| (BBB2) | 6.04E+14 | 0.0 | 0.674 | 0.645 | 11.339 | 11.339 | 0.000 |
|        | 6.64E+14 | 580.0 | 0.857 | 0.812 | 12.299 | 12.299 | 0.495 |
|        | 7.60E+14 | 861.5 | 1.146 | 1.071 | 13.555 | 13.555 | 0.821 |
|        | 8.20E+14 | 963.5 | 1.308 | 1.212 | 14.131 | 14.131 | 0.941 |
|        | 9.00E+14 | 1063.3 | 1.499 | 1.376 | 14.720 | 14.720 | 1.036 |
|        | 1.00E+15 | 1155.8 | 1.699 | 1.544 | 15.248 | 15.248 | 1.084 |
| (BPAL12) | 7.57E+14 | 0.0 | 0.911 | 0.860 | 12.476 | 12.476 | 0.000 |
|          | 8.00E+14 | 353.6 | 0.975 | 0.917 | 12.739 | 12.739 | 0.256 |
|          | 8.73E+14 | 580.0 | 1.082 | 1.013 | 13.202 | 13.202 | 0.442 |
|          | 1.28E+15 | 1047.3 | 1.503 | 1.378 | 14.647 | 14.647 | 0.820 |
|          | 1.36E+15 | 1094.1 | 1.555 | 1.421 | 14.782 | 14.782 | 0.836 |
|          | 1.44E+15 | 1134.7 | 1.599 | 1.458 | 14.891 | 14.891 | 0.845 |

1.5) M$_\odot$, (0.6–2.1) M$_\odot$ and (1.4–2.2) M$_\odot$ respectively. For lower values of the highest QPO frequency, this range will widen making the constraints even less firm, unlike in case (a). However, if in any system a QPO frequency is discovered which lies above the maximum allowed $\nu_K$ (the peaks in fig. 1) for the EOS, then the corresponding EOS can certainly be ruled out.

*Clearly, the constraints placed on neutron star masses (as well as those on the EOS) are dependent on the model assumed for the generation of kHz QPOs.*

In Tables 1 and 2 we have listed the values of the neutron star central density ($\rho_c$), spin frequency ($\nu_S$), the baryonic mass ($M_0$), the gravitational mass ($M_G$), the radius ($R$), radius of the innermost allowed orbit ($r_{iao}$). For the sake of a comparison with the calculations that are based on the 'slow' rotation approximation, we have also listed in Table 1 $r_{ms,\mathrm{HT}}$ and $\nu_{K,\mathrm{HT}}$ for several values of the dimensionless specific angular momentum $j$ ($\equiv J/M_G^2$, $J$ being the angular momentum of the 'slowly' rotating neutron star). The 'slow' rotation approximation is based on the Hartle & Thorne (HT) formalism (Hartle & Thorne 1968), which treats rotation as a perturbation to the spherically symmetric space–time. This formalism gives the following analytical expressions for the radius of the innermost marginally stable orbit and the corresponding Keplerian frequency (Kluźniak & Wagoner 1985; Kluźniak, Michelson & Wagoner 1990):

$$r_{ms,\text{HT}} = 6M_G(1-(2/3)^{3/2}j) \qquad (1)$$

$$\nu_{K,\text{HT}} = \frac{1}{2\pi}\left[1-\left(\frac{r}{M_G}\right)^{-3/2}j\right]\left(\frac{M_G}{r^3}\right)^{1/2} \qquad (2)$$

Here $c = 1 = G$. The HT formalism is valid only for those values of the stellar angular velocity that are small in comparison to the centrifugal break–up angular velocity.

The quantities listed in the tables have been calculated by choosing a combination of $\rho_c$ and $\nu_S$ so as to obtain a value of $\nu_K$ equal to 1220 Hz. Unlike in Fig. 1, for the results that are presented in the tables, no specific value of $\nu_S$ was chosen but instead we chose a fixed value for $\nu_K = 1220$ Hz, while exploring various values of $\nu_S$. Table 1 represents Solution I corresponding to the case $r_{iao} = r_{ms}$ and Table 2 represents Solution II for which $r_{iao} = R$. Only those EOS models which allow Solution I are displayed in Table 1. For EOS (BBB2), Solution I is allowed only for very slowly rotating configurations with $\nu_S < 100$ Hz and hence there is only one entry against this EOS model in Table 1. For the sake of illustration, we include in the Tables 1 and 2 results for neutron star rotation frequency of 580 Hz as inferred for the source 4U 1636–53 (Zhang et al. 1997b) for all those configurations for which such solutions exist. Table 1 shows that though $r_{ms,\text{HT}}$ varies at most by 10% from $r_{iao}$, the differences are substantially larger for the corresponding values of $\nu_{K,\text{HT}}$ particularly for large rotation rates.

It is interesting to ask what the maximum value of $\nu_K$ would be for a given EOS model, because this value represents the boundary between Solution I and II. In Table 3 we list the maximum values of $\nu_K$ along with the corresponding $M_G$ and $R$ for different chosen values of $\nu_S$, namely, 0, 200 and 580 Hz. It is clear from this table that the softer the equation of state, the higher the value of the maximum of $\nu_K$. It may be relevant to note that the EOS model SBD has a maximum $\nu_K$ value of 1259 Hz for $\nu_S = 580$ Hz and a corresponding mass of 2.02 M$_\odot$.

## 3 CONCLUSIONS

The results of the present study can be summarized as follows. We find that the identification of the maximum observed (so far) QPO frequency (1220 Hz) with $\nu_K$ implies the following constraints on the mass of the neutron star and the EOS:

(a) If Solution I is the preferred scenario for kHz QPOs, then with the additional assumption that the highest observed QPO frequencies are produced only at $r = r_{ms}$, soft EOS models for neutron star interiors would be disfavoured.

(b) If the highest observed QPO frequencies are produced at $r = r_{iao}$, another set of constraints, possessing lower values of mass can be deduced. Together with the constraints obtained from case (a) these results limit the neutron star mass values to be in either of the two ranges: (0.6–1.8) M$_\odot$ or (1.8–2.2) M$_\odot$.

(c) If the highest observed QPO frequencies are generated at the inner edge of the accretion disk, which may be located outside the innermost "allowed" orbit, the constraints implied on the neutron star mass will not be firm ones and this value may lie within the range (0.6–2.2) M$_\odot$. Furthermore, *all EOS models allow $\nu_K \geq 1220$ Hz values.*

The above results may be considered to be of sufficient generality as we have considered EOS models ranging from very stiff to very soft. We emphasise that the above conclusions, which constitute a set of constraints, are a consequence of the inclusion of general relativistic rotational effects and the non–restrictive assumption $\nu_K \geq 1220$ Hz. It is relevant to mention here that the constraints on the EOS and the neutron star mass not including the rotational effects in general relativity and assuming $R < r_{ms}$, would be somewhat more firm but *less general*, discounting their utility. Fig. 1 and Table 3 show that the maximum allowed $\nu_K$ for all EOS models that we have considered exceed 1220 Hz. We believe, therefore, that determination of the maximum QPO frequency and spin rate of the neutron stars among a substantially large sample of low–mass X–ray binaries at various evolutionary stages in future observations will provide more meaningful and general constraints on neutron star masses and their EOS.

*Note added*: After the submission of this paper, a work on a similar theme appeared in the pre–print archives (Miller, Lamb & Cook 1998). The results reinforce our conclusions.

## ACKNOWLEDGEMENTS

The authors thank an anonymous referee for helpful comments which improved the text presentation.

This paper has been produced using the Royal Astronomical Society/Blackwell Science LaTeX style file.

**Table 3.** Maximum $\nu_K$ attainable for each EOS for three values of spin frequency $\nu_S$ of the neutron star. Also listed are the corresponding values of gravitational mass $M_G$ and radius $R$.

| EOS | $\nu_S = 0$ (Hz) | | | $\nu_S = 200$ (Hz) | | | $\nu_S = 580$ (Hz) | | |
|---|---|---|---|---|---|---|---|---|---|
| | $\nu_K$ (Hz) | $M_G$ (M$_\odot$) | $R$ (km) | $\nu_K$ (Hz) | $M_G$ (M$_\odot$) | $R$ (km) | $\nu_K$ (Hz) | $M_G$ (M$_\odot$) | $R$ (km) |
| (C) | 1783 | 1.224 | 10.898 | 1819 | 1.284 | 10.884 | 1817 | 1.372 | 11.122 |
| (N) | 1560 | 1.330 | 12.250 | 1606 | 1.468 | 12.364 | 1584 | 1.619 | 12.857 |
| (UU) | 1743 | 1.244 | 11.122 | 1774 | 1.320 | 11.170 | 1769 | 1.439 | 11.497 |
| (SBD) | 1303 | 1.692 | 15.008 | 1306 | 1.800 | 15.208 | 1259 | 2.021 | 16.235 |
| (BBB2) | 1732 | 1.258 | 11.218 | 1760 | 1.317 | 11.226 | 1770 | 1.436 | 11.494 |
| (BPAL12) | 1752 | 1.254 | 11.106 | 1800 | 1.293 | 10.989 | 1840 | 1.373 | 10.974 |